\documentclass{article}
\usepackage{amsfonts}
\usepackage{amssymb}
\usepackage{graphics}
\def\doeack{This work was supported in part by the Department of 
Energy, Nuclear Physics Division, under contract DE-FG02-86ER40286}
\def\beq{\begin{equation}}
\def\eeq{\end{equation}}
\def\beqarray{\begin{eqnarray}}
\def\eeqarray{\end{eqnarray}}

\title{Scattering Calculations with Wavelets}
\author{B. M. Kessler, G. L. Payne, W. N. Polyzou\thanks{\doeack}\\  
The University of Iowa, Iowa City, IA 52242, USA}

\begin{document}

\maketitle
\begin{abstract}

We show that the use of wavelet bases for solving the momentum-space
scattering integral equation leads to sparse matrices which can
simplify the solution.  Wavelet bases are applied to calculate the
K-matrix for nucleon-nucleon scattering with the s-wave Malfliet-Tjon
V potential.  We introduce a new method, which uses special properties
of the wavelets, for evaluating the singular part of the integral.
Analysis of this test problem indicates that a significant reduction
in computational size can be achieved for realistic few-body
scattering problems.

\end{abstract}  

\section{Introduction}

In this paper we show that bases of compactly supported wavelets can
lead to a significant reduction in the size of the matrices used to
solve few-body momentum-space scattering integral equations.  When the
kernel of the integral equation is expanded in a wavelet basis, the
resulting matrix can be expressed as the sum of a sparse matrix and a
matrix with small norm.  Ignoring the small matrix leads to an
accurate approximate solution.

Our interest in wavelet bases is motivated by the successful
application of wavelets to data compression problems.  The
coefficients of a wavelet basis provide a compact representation for
storing information that has smooth structures on different scales.
Wavelets are used for data compression in the FBI fingerprint archive
\cite{fbi} and JPEG files \cite{jpeg}.  The property that makes
wavelets useful for data compression suggests that they may also lead
to sparse matrix representations of the kernels of the momentum-space
scattering integral equations.  Conventional methods lead to 
dense matrix representations of these equations; the computation 
of the solution of these equations for realistic models requires 
solving large systems of linear equations.

While the use of wavelet bases lead to sparse matrices in both
momentum space and configuration space, the resulting reduction in the
size of the equations in momentum space is significant, because the
momentum-space kernels are represented by full matrices.  These
advantages are important for relativistic scattering problems, which
are naturally formulated in momentum space.

There are many different types of wavelets that are used in specific
applications.  In this paper we use the compactly supported wavelets
introduced by Daubechies \cite{Daubechies}\cite{Daubechies2} in 1988
to evaluate the K-matrix for two spinless nucleons interacting with
an s-wave Malfliet-Tjon V potential \cite{Malfliet}.  
Nucleon-nucleon scattering with the Malfliet-Tjon V potential provides
a good test of numerical methods because the on-shell energy and
strength of the attractive and repulsive parts of the potential lead
to a problem with three different momentum scales. 

The Daubechies wavelets were the first non-trivial set of orthonormal
wavelets of compact support discovered, and they have a number of
properties that make them useful for numerical applications.  However,
the Daubechies wavelet basis functions involve a significant amount of
structure that can make it difficult to appreciate some of these
advantages.  Some of the properties that we have found useful are
briefly discussed below.  They will be explained in more detail in 
section two.

In applications, there are two bases, called the wavelet basis and the
scaling basis. They are related by an orthogonal transformation,
called the wavelet transform.  This orthogonal transformation can be
implemented efficiently using wavelet filters \cite{numrec}.
The scaling basis is constructed from a single function, which is
called the scaling function or father function. The scaling basis
consists of discrete translates of a dyadic (power of 2) scale
transformed copy of the original scaling function.  The scale, which
is called the fine scale, is determined by the application.  The
wavelet basis is also constructed from a single function, called the
mother wavelet.  The wavelet basis consists of discrete translations
of dyadic scale-transformed copies of the mother wavelet function.
These basis functions are called wavelets.  This basis includes
wavelets on all dyadic scales between the fine scale and a coarse
scale, which is also dictated by the application.  The wavelet basis
is completed by translates of the scaling function on the coarse
scale.
 
The expansion coefficients for a function in the scaling basis are
equivalent to weighted averages of the function over the support of
the scaling basis functions.  The wavelet basis provides an efficient
representation to store this information on multiple scales.  The
principal advantages of the wavelet method are best explained in terms
of properties of these two complementary sets of basis functions.

\begin{itemize}

\item[1.] The wavelets and scaling basis functions are orthonormal
functions with compact support.  This reduces global errors when these
bases are used to represent local structures.

\item[2.] The wavelet and scaling basis functions are related to a single
``mother (resp. father) function'' by unit translations and dyadic
scale changes.  This simplifies the computation of the basis
functions.

\item[3.] Finite linear combinations of the scaling basis functions
can locally represent low-degree polynomials.  This means that kernels
which can be locally approximated by low-degree polynomials on the
fine scale can be accurately represented by expansions in a scaling
basis.  This is the defining property of the Daubechies wavelets.

\item[4.] Wavelet basis functions are orthogonal to low-degree
polynomials.  This is the reason why the kernel in a wavelet basis is
sparse.  If the kernel can be locally approximated by a polynomial on
a sufficiently small region, the matrix elements corresponding to the
wavelets with support on that region will be small or zero. In this 
representation the kernel is accurately represented by a small number of 
wavelet basis functions on a large scale. Since the wavelet basis is 
equivalent to the scaling basis,  it follows that the wavelet basis 
can accurately represent kernels that can be locally approximated by 
low-degree polynomials on the fine scale by sparse matrices.

\item[5.] All moments of wavelets or scaling basis functions can be
computed exactly and efficiently.  These moments are essential for
performing the integrals needed for a typical calculation.

\item[6.] There is a one-point quadrature rule which integrates the
product of second degree local polynomials and scaling basis functions
exactly.  In our applications, the one-point quadrature rule reduces
the Galerkin method, which constructs an approximate matrix equation
by projecting the original equation on a subspace spanned by a set of
basis functions, to the collocation method, which constructs an
approximate matrix equation by demanding that the equation hold on a
set of collocation points.  The resulting systems of equations share
the advantages of both methods.

\item[7.] The wavelet transform provides a means to automatically
determine the important basis functions.  Specifically, it is enough to
set all matrix elements of the transformed kernel smaller than 
a specified minimal size to zero, which yields a sparse matrix.

\item[8.] Integrals over the scattering cut can be computed accurately
and efficiently using the scaling equation.  These integrals can
be expressed in terms of non-singular integrals that can be evaluated
in terms of the known moments.

\item[9.] The integral equation can be solved in the wavelet basis
without computing a wavelet or a scaling basis function.  This technique 
exploits many of the properties of the Daubechies wavelets and it saves
the computation time involved in evaluating the basis functions. 

\end{itemize}
These features will be explained in more detail in the body of the 
paper.

Given the many useful properties of Daubechies wavelets, it is natural
to ask why they have not been extensively used to treat scattering
problems.  We speculate that part of the problem is the structure of
the wavelets.  Wavelets are different than the smooth functions that
are normally used in numerical analysis.  They are finite linear
combinations of solutions of a scaling equation, which relates a
function on one scale to a linear combination of translates of the
same function on a smaller scale.  The solution of this scaling
equation has a self-similar or fractal structure, which repeats itself
on arbitrarily small scales.  Functions with fractal structure are not
amenable to numerical methods that exploit the smoothness of functions
on a sufficiently small scale.  Our experience is that typical
adaptive quadrature methods are very inefficient when applied to
wavelets.  We found this to be especially true for integration over
the scattering singularity.

While the fractal structure of wavelets limits the utility of 
conventional numerical methods,  the scaling equation provides an
alternative tool, which when properly exploited,  can be used to   
overcome {\it all} of the difficulties that arise with conventional methods.
An important message in this paper is that success is obtained by 
embracing the scaling equation.  

In preparing this manuscript we found a number of references on
wavelets to be useful in understanding how to apply wavelet bases to
solve integral equations.  These include
\cite{fbi}\cite{Daubechies}\cite{Daubechies2}\cite{numrec}\cite{Resnik}\cite{Strang}\cite{Shann1}\cite{Shann2}\cite{Sweldens}
and \cite{Kaiser}.  In addition, more details can be found in our web
notes \cite{web}, which provide a detailed summary of many of the
things that we learned from these references.  A key element that we
believe is new in this paper is the treatment of the singular
integral, which has proven to be both stable and accurate.  This is 
also discussed in detail in these notes.

\section{Wavelets} 

In this section we define Daubechies' initial family of
compactly supported orthogonal functions \cite{Daubechies2}.  The
defining property of Daubechies' wavelets is that they 
are orthogonal to low-degree polynomials.

As mentioned earlier, the basis functions consist of discrete
translations and scale transforms of two functions, called the scaling
function, which we denote by $\phi (x)$, and mother wavelet function,
which we denote by $\psi(x)$.  An approximation space has two bases
related by a fast orthogonal transformation, called the wavelet
transform. One basis consists of translates of the scaling function on
a fine scale. The second basis consists of translates of the scaling
function on a coarse scale and the wavelets on all dyadic scales
between the fine and coarse scale.  The first representation will lead
to a dense matrix representation of the integral equations while the
second results in a sparse matrix representation.  The choice of fine scale
defines the approximation space; it is determined by the finest
structure in the driving term and the kernel of the integral equation.

We begin by discussing the scaling function, $\phi (x)$, which has
compact support and integrates to a value of one. 
To construct the scaling function we utilize two operations, dilation
and translation. They are implemented by the unitary operators:
\beq
\hat{D} \phi(x) = {1 \over \sqrt{2}}\phi(x/2)
\label{eq:BA}
\eeq
which is a dyadic scale transformation and
\beq
\hat{T} \phi (x) = \phi (x - 1).
\label{eq:BB}
\eeq
which is a unit translation.
The scaling function $\phi (x)$ is the solution of the scaling equation
\beq
\hat{D} \phi (x) = \sum_{l=0}^{2K-1} h_l \hat{T}^l \phi (x)
\label{eq:BC}
\eeq
with normalization
\beq
\int \phi (x) dx =1.
\label{eq:BD}
\eeq
Solutions, $\phi (x)$, of the scaling equation with finite $K$ are 
of special interest because these solutions have compact support 
in the interval $[0,2K-1]$ \cite{web}. 

The scaling equation is a linear renormalization group equation that
relates the scaling function on a coarse scale to linear combinations
of the scaling function on a fine scale.  The coefficients $h_l$,
called the scaling coefficients, are numerical coefficients which
define the type of scaling function.  For most computations, a
knowledge of the $h_l$'s is all that is needed.  In some wavelet 
literature the normalization convention for the scaling coefficients, $h_l$,
differs by a factor of $\sqrt{2}$

The scaling basis functions, $\phi_{j,k}(x)$, are defined by 
taking $j$ dilations and $k$ translations of the scaling 
function:
\beq
\phi_{j,k} ( x ) := \hat{D}^j \hat{T}^k \phi (x) = 
2^{-j/2} \phi \left ({x\over 2^j}-k \right ) . 
\label{eq:BE}
\eeq
Our convention, which follows refs. \cite{Daubechies}\cite{Daubechies2},
is that $-\infty < j < \infty$, with decreasing $j$ corresponding to finer 
scales and increasing $j$ corresponds to coarser scales; some 
wavelet references use the opposite convention, with increasing $j$ 
corresponding to finer scales. 
The scaling function $\phi (x)$ is defined so that the translates of 
the $\phi_{j,k}(x)$ on a fixed scale $j$ are orthonormal;
\beq
(\phi_{j,k},\phi_{j,l}) = \delta_{kl}. 
\label{eq:BF}
\eeq
Self-consistency of the scaling equation (\ref{eq:BC}) and the orthonormality 
requirement (\ref{eq:BF}) constrain the coefficients $h_l$ so that 
\beq
\sum_{l=0}^{2K-1} h_l = \sqrt{2}, \qquad
\sum_{l=0}^{2K-1} h_l h_{l-2n} = \delta_{n0}.
\label{eq:BG}
\eeq
For $K=1$ these conditions fix the two coefficient $h_0$ and $h_1$ 
uniquely. The $K=1$ solution is called the Haar \cite{Daubechies}\cite{haar} 
scaling function.  For larger values of $K$, additional constraints are 
needed to determine the coefficients $h_l$.  These constraints will be 
discussed later in the paper.  In all cases the, the number of $h_l$'s 
must be even and the support of the scaling function can be shown \cite{web} 
to be in the interval $[0,2K-1]$.  The support of $\phi_{j,l}(x)$ has 
width $2^{j}(2K-1)$. That the scaling equation has non-trivial solutions 
with compact support was not known before Daubechies seminal 
paper \cite{Daubechies}. 

The subspace of $L^2({\Bbb R} )$ spanned by $\phi_{j,k}(x)$ for fixed $j$ is
denoted by ${\cal V}_j$. This space is the approximation
space in our calculations.  The parameter $j$ defines the scale of
these functions, and the size of the support of 
$\phi_{j,k}(x)$.   The scale gets finer as $j$ decreases, and $j$ 
is determined by the finest scale of the problem.  Two
important properties of the different subspaces ${\cal V}_j$ are that
\beq
{\cal V}_{j-1} \supset {\cal V}_{j}
\label{eq:BH}
\eeq
and that ${\cal V}_{j}$ becomes dense in $L^2({\Bbb R})$ as $j \to -\infty$. 
Equation (\ref{eq:BH}) is a consequence of the scaling equation. 
 
The mother wavelet $\psi (x)$ is defined to be orthogonal to the
scaling function and its translates.  The mother wavelet is associated
with changes in the represented function.  It can be visualized as a
small wave that integrates to zero.  To define the mother wavelet
recall Eq. (\ref{eq:BH}).  For each $j$ define ${\cal W}_{j}$ to be the
orthogonal complement of ${\cal V}_{j}$ in ${\cal V}_{j-1}$:
\beq
{\cal V}_{j-1} = {\cal W}_j \oplus {\cal V}_{j} .
\label{eq:BJ}
\eeq
Orthonormal basis functions of ${\cal W}_{j}$ are linear combinations
of the scaling basis functions of ${\cal V}_{j-1}$.  The different basis
functions of ${\cal W}_j$ are discrete translations of a single function. 
The mother wavelet $\psi (x)$ is the element of ${\cal W}_0$ defined by
the following linear combination of the scaling basis functions on
${\cal V}_{-1}$:
\beq
\psi (x):= \hat{D}^{-1} \sum_{l=0}^{2K-1} g_k \hat{T}^l \phi (x)
\label{eq:BK}
\eeq
where the coefficients 
\beq
g_l := (-1)^l h_{2K-1-l}
\label{eq:BL}
\eeq
are fixed by the orthonormality conditions: 
\beq
(\hat{T}^l \psi , \hat{T}^m \psi ) = \delta_{lm}, \qquad
(\hat{T}^l \psi, \hat{T}^m \phi) =0.
\label{eq:BM}
\eeq
The wavelets $\psi_{0,l}(x) :=\hat{T}^l \psi (x)$ span ${\cal W}_0$, while
\beq
\psi_{j,l} (x) := \hat{D}^j \hat{T}^l \psi (x). 
\label{eq:BN}
\eeq
is an orthonormal basis for ${\cal W}_j$.  Note that the number of 
basis functions doubles as $j$ is decreased by one.  The support of 
$\psi_{j,l} (x)$  is identical to the support of $\phi_{j,l} (x)$,
which is 
\[
\mbox{supp} (\phi_{j,l}(x) )=
\mbox{supp} (\psi_{j,l}(x) )
= [2^j l, 2^j(l+2K-1)].
\]
For a large negative value of $j$ we can iterate (\ref{eq:BJ})  
to decompose ${\cal V}_j$ on a fine scale into a direct sum of ${\cal W}_k$'s
on coarser scales:
\beq
{\cal V}_j = {\cal W}_{j+1}  \oplus  {\cal W}_{j+2} \oplus \cdots \oplus 
{\cal W}_{j+m} \oplus  {\cal V}_{j+m}.
\label{eq:BO}
\eeq
In wavelet terminology this is called the multi-resolution
decomposition of ${\cal V}_j$.  The right- and left-hand sides of
Eq. (\ref{eq:BO}) define two bases for the space ${\cal V}_j$.  The
basis on the left consists of just the scaling basis functions
$\phi_{j,l}(x)$ on the fine scale ``$j$''.  The basis on the right
consists of wavelets, $\psi_{k,l}(x)$, on all scales $k$ from the
finest $(j+1)$ to the coarsest $(j+m)$ and the scaling basis functions
$\phi_{j+m,l}(x)$ on the coarsest scale, $(j+m)$. Note that unlike the
spaces ${\cal V}_{j}$, the spaces ${\cal W}_j$ on different scales are
orthogonal.  This leads to the following alternative representations
of functions in ${\cal V}_j$:
\beq
f(x) = \sum_l c_l \phi_{j,l} (x) = 
\sum_l d_l \phi_{j+m,l}(x)+ \sum_{k=j+1}^{j+m} \sum_{l} d_{k,l} \psi_{k,l} (x)
\label{eq:BP}
\eeq
where the expansion coefficients are
\beq
c_l := (\phi_{j,l},f), \quad d_l := (\phi_{j+m,l},f) \quad
d_{k,l}:= (\psi_{k,l},f) .
\label{eq:BQ}
\eeq
It follows from Eqs. (\ref{eq:BO}) and (\ref{eq:BP}) that there is an
orthogonal transformation relating the coefficients $c_l$ to $d_l$ and
$d_{k,l}$.

When we approximate a function $f(x)$ as a linear combination 
the scaling basis functions, $\phi_{j,l} (x)$, and the original
function $f(x)$ is approximately constant, $f(x)=f_{l}$, on the support of
$\phi_{j,l}(x)$, the expansion coefficients, $c_l =
(f,\phi_{j,l})\approx 2^{j/2} f_l$, are up to a scale factor, the value
of the function at any point in the support of $\phi_{j,l}(x)$.  
For a continuous function $f(x)$, and a sufficiently fine
scale $j$, it follows that, up to a scale factor, the 
coefficients $c_l$ track the value of $f$ in the sense:
\beq
c_l \approx  2^{j/2} f(2^j l ).
\label{eq:BQA}
\eeq
The coefficients 
$c_l$ form a very inefficient approximate representation of the 
function $f(x)$; while the coefficients $d_l$ and $d_{k,l}$ form an
efficient representation, since many of the $d_{k,l}$ will be small
and can be neglected without introducing substantial errors. 

The Daubechies wavelets of order K are defined by the conditions that 
the mother wavelet satisfy
\beq
\int x^k \psi (x) dx = 0, \quad 0 \leq k \leq K-1 .
\label{eq:BR}
\eeq
Using Eq. (\ref{eq:BR}) gives the additional 
constraints needed to determine the scaling coefficients $h_l$:
\beq
0 = \sum_{l=0}^{2K-1} g_l l^k = 
\sum_{l=0}^{2K-1} (-1)^l h_{2K-1-l} l^k, \quad 0 \leq k \leq K-1 .
\label{eq:BS}
\eeq
These equations, along with Eq. (\ref{eq:BG}), determine the
Daubechies scaling coefficients, $h_l$, of order $K$ up to a reversal
in the order of the coefficients.  Note that the second equation in
(\ref{eq:BG}) is actually several equations.  There is a different 
equation for each value of $n$.  For the case $K=1,2,3$ these equations can be
solved analytically. The solutions are given in Table 1 \cite{numrec}.
\begin{table} 
{\bf Table 1: Scaling Coefficients } \\[1.0ex]
\begin{tabular}{|l|l|l|l|}
\hline
$h_l$ & K=1 & K=2 & K=3   \\
\hline					      		      
$h_0$ &$1 / \sqrt{2}$  & $(1+\sqrt{3})/4\sqrt{2}$ &$(1+\sqrt{10}+\sqrt{5+2\sqrt{10}})/16\sqrt{2}$ \\
$h_1$ &$1 / \sqrt{2}$  & $(3+\sqrt{3})/4\sqrt{2}$ & $(5+\sqrt{10}+3\sqrt{5+2\sqrt{10}})/16\sqrt{2}$ \\
$h_2$ &$0$  & $(3-\sqrt{3})/4\sqrt{2}$ & $(10-2\sqrt{10}+2\sqrt{5+2\sqrt{10}})/16\sqrt{2}$ \\
$h_3$ &$0$  & $(1-\sqrt{3})/4\sqrt{2}$ & $ (10-2\sqrt{10}-2\sqrt{5+2\sqrt{10}})/16\sqrt{2} $ \\
$h_4$ &$0$  & $0$ & $(5+\sqrt{10}-3\sqrt{5+2\sqrt{10}})/16\sqrt{2}$ \\
$h_5$ &$0$  & $0$ & $(1+\sqrt{10}-\sqrt{5+2\sqrt{10}})/16\sqrt{2}$ \\
\hline
\end{tabular}
\end{table}
For larger values of $K$ these equations can be solved numerically
and tables of the scaling 
coefficients $h_l$ for $K>3$ can be 
found in the literature \cite{Daubechies}\cite{Daubechies2}. 
It is straightforward to show that Eq. (\ref{eq:BR})
holds on all scales: 
\beq 
\int \psi_{j,l}(x) x^k dx = 0, \quad 0 \leq k \leq K-1 . 
\label{eq:BT}
\eeq 
Plots of the $K=2$ and $K=3$ scaling function $\phi (x)$ and mother
wavelet $\psi (x)$ are shown in Figures 1 and 2.  The functions
with $K=2$ have support on the interval $[0,3]$ while the $K=3$
functions have support on $[0,5]$.  The scaling function and mother
wavelet can be distinguished by the conditions that the integral of
the scaling function is one while the integral of the mother wavelet
is zero.  In both figures the compact support and fractal nature of
these functions are apparent.

One of the key features of the Daubechies wavelets follows 
from Eq. (\ref{eq:BT}).  Note that in the limit $j \to -\infty$
the expansion (\ref{eq:BP}) becomes exact.  If we expand 
\beq 
x^k = \sum_l d_l \phi_{j,l}(x)+ \sum_{k=-\infty}^{j} \sum_{l} d_{k,l} 
\psi_{k,l} (x),
\label{eq:BU}
\eeq 
then the condition (\ref{eq:BT}) gives $d_{k,l}=0$ for $k<K$, or 
\beq 
x^k = \sum_l d_l \phi_{j,l}(x) .
\label{eq:BV}
\eeq
While $x^k$ is not a square-integrable function, at any point $x$ 
the support conditions imply that the above sum has no more than 
$2K-1$ non-vanishing terms.  It follows that on any compact subset 
this sum is {\it pointwise} identical to $x^k$.  Note that
this is true for any scale $j$.  This explains why the scaling function 
basis and the equivalent wavelet basis can be used to locally represent 
polynomials.

Eq. (\ref{eq:BT}) explains why the second basis in Eq. (\ref{eq:BP})
is so efficient.  If the function $f(x)$ can be locally approximated 
by a polynomial of degree $K$ on the support of $\psi_{j,l}(x)$, then 
the coefficients $d_{k,l}:= (\psi_{k,l},f) \approx 0$.
 
Although the wavelets with higher values of $K$ have some nicer
properties, their support and the number of non-zero basis functions
at any given point $x$ increases with $K$.  Thus, there is a point
where the additional smoothness \cite{Kaiser} of a higher-order
wavelet is balanced against the cost of manipulating functions with
larger supports and more scaling coefficients, $h_l$.  For our application,
we believe that the Daubechies $K=3$ wavelet is the optimal choice to
balance accuracy with computational speed.

The important observation about the Daubechies wavelets is that once
the scaling coefficients $h_l$ are determined, all of the calculations
can be reduced to manipulations involving just these coefficients.
This is done by making judicious use of the scaling equation to derive
linear relationships for the quantities that we desire.  This method
is used to calculate all moments of the scaling and wavelet functions
exactly and to construct quadrature rules.  It can also be used to
compute both $\psi (x)$ and $\phi (x)$ exactly at all dyadic points.
Furthermore, the scaling equation leads to the transformation that we
need to transform between the scaling basis function representation
and the wavelet basis representation.  The scaling equation is also used
to accurately evaluate the integral of the product of the scaling
basis functions and the scattering singularity.  Specific applications
of the scaling equation that are used to solve for the $K$-matrix are
outlined below.

\noindent{\bf A. Computation of $\phi (x)$ and $\psi(x)$:}

To calculate $\phi (x)$ we  use the scaling equation in the form
\beq
\phi (x) = \sum_{l=0}^{2K-1} \sqrt{2} h_l \phi (2x-l). 
\label{eq:BW}
\eeq
Setting $x$ to integer values $n$ gives linear relations among the 
non-zero $\phi (n)$'s at integer points:
\beq
\phi (n) = \sum_{l=0}^{2K-1} \sqrt{2} h_l \phi (2n-l). 
\label{eq:BX}
\eeq
These homogeneous equations can be supplemented with the inhomogeneous 
equation
\beq
1 = \sum_n \phi (n)  
\label{eq:BY} 
\eeq
to get a linear system of equations for $\phi (x)$ at all integer points.
Eq. (\ref{eq:BY}) follows by using the normalization condition 
(\ref{eq:BD}) to compute the expansion coefficients (\ref{eq:BP}) 
for $f(x)=1$, and evaluating the result at an integer value of $x$.  

Given the values $\phi (n)$, recursive use of the scaling equation
(\ref{eq:BW}) or (\ref{eq:BK}) gives exact values of $\phi (x)$ or
$\psi (x)$ at all dyadic rationals.  Since these functions are
continuous \cite{Daubechies2} for $K>1$, and the dyadic rationals are
dense in the real numbers, $\phi (x)$ and $\psi (x)$ can be calculated
at all points by continuity.  While this method is exact at dyadic
points, it is computationally intensive; however, the method discussed
in this paper does not require the evaluation of the wavelets or the
scaling basis functions.

\noindent{\bf B. Moments:}

Moments of the scaling function are defined by 
\beq
<x^k>_{jl} \, := \int \phi_{j,l}  (x) x^k dx  .
\label{eq:BZ}
\eeq
Using
\beq
(\hat{D}^j\hat{T}^l\phi , x^k)= (\phi , \hat{T}^{-l} \hat{D}^{-j} x^k ) 
\label{eq:BAA}
\eeq
these moments can be expressed in terms of the moments $<x^k>_{00}$. 
Using the unitarity of the dilation operator and the scaling equation in 
\beq
<x^k>_{00} = (\phi , x^k) = (\hat{D}\phi , \hat{D}x^k)
\label{eq:BAB}
\eeq
gives
\beq
<x^k>_{00} = {1 \over 2^k} \sum_l {h_l \over \sqrt{2}} \sum_{m=0}^k
{ k! \over m! (k-m)!} l^{k-m} <x^m>_{00} .
\label{eq:BAC}
\eeq
This equation can be used to express the higher moments in terms of the 
lower moments
\[
<x^k>_{00} =
\]
\beq
{1 \over 2^{k}-1} \sum_l {h_l \over \sqrt{2}} \sum_{m=0}^{k-1}
{ k! \over m! (k-m)!} l^{k-m} <x^m>_{00} 
\label{eq:BAD}
\eeq
which can be used with the normalization condition (\ref{eq:BD}) to 
recursively compute moments of any order.  Given the moments of 
the scaling function, Eq. (\ref{eq:BK}) can be used to compute moments 
of the wavelets.

\noindent{\bf C. Quadrature Rules:}

Given a set of moments it is always possible to use them to construct 
quadrature rules.  In most elementary applications a set of 
quadrature points $\{x_k\}$ is given.  The system of linear 
equations 
\beq
\sum_k (x_k )^n w_k  = <x^n>_{00} 
\label{eq:BADX} 
\eeq   
is used to solve for the weights $w_k$ in terms of the 
$<x^n>_{00}$ moments.  For applications involving a scaling basis
function on a fine scale only a small number of quadrature points 
are needed.

In this paper most of the integrals are performed using 
a simple and effective one-point quadrature
rule that follows from the identity
\beq
<x^2>_{00} = <x^1>_{00}^2 = \left ({1 \over 
\sqrt{2}} \sum_{l=1}^{2K-1} l h_l \right )^2,
\label{eq:BAE}
\eeq
which holds for the moments of the Daubechies $K>1$ scaling basis functions.  
In this case one obtains 
\beq
\int \phi (x) p(x) dx = p(<x^1>_{00}) 
\qquad \mbox{for} \qquad p(x) =a +bx +cx^2 
\label{eq:BAF}
\eeq
which provides a one-point quadrature rule\cite{Sweldens}, 
with point $x=<x^1>_{00}$
and weight $w=1$,  
that integrates the 
product of the polynomial and the scaling function exactly.
Translates and dyadic scale transformations of these moments 
give one-point quadrature 
rules for each of the $\phi_{j,l}(x)$.  In applications we use 
points $<x^1>_{jl}$ and weights $w_{jl}$ for $\phi_{j,l}(x)$ on a 
small scale $j$:
\beq
< x^1>_{jl} = 2^{j} (<x^1>_{00} + l) \quad w_{jl} = 2^{j/2}
\label{eq:BAG}
\eeq
This quadrature rule is used on the finest scale with the Daubechies
K=2,3 scaling basis functions, which can represent local
polynomials of degree one and two exactly.  For Daubechies wavelets 
with larger K-values, which can locally represent polynomials of degree 
$K-1$,  multi-point quadratures rules can be constructed 
which exactly integrate the product of the scaling function with 
degree $K-1$ polynomials. 

\noindent{\bf D. Partial Moments:} 

In solving integral equations on a finite or semi-infinite interval, 
it is necessary to calculate integrals where an end-point of the 
interval is in the support of a scaling basis function 
\cite{Shann1}\cite{Shann2}.  By using the scaling and dilation 
operators these integrals are simply related to the following 
integrals:
\beq
I^{k}_{m+}:= \int_{0}^{\infty} \hat{T}^m \phi (x) x^k dx,
\qquad
I^{k}_{m-}:= \int_{-\infty}^{0} \hat{T}^m \phi (x) x^k dx
\label{eq:BAH}
\eeq
and
\beq
I^{0}_{mn+}:= \int_{0}^{\infty} \hat{T}^m \phi (x) \hat{T}^n \phi (x)  dx,
\qquad
I^{0}_{mn-}:= \int_{-\infty}^{0} \hat{T}^m \phi (x) \hat{T}^n \phi (x)  dx .
\label{eq:BAJ}
\eeq
To calculate these quantities one can use the scaling equation in the form 
(\ref{eq:BW}).  This leads to the linear relations 
\beq
I^k_{m+} = 2^{-k-1/2}  \sum_{l=0}^{2K-1} h_l I^k_{2m+l,+}
\label{eq:BAM}
\eeq
\beq
I_{mn+} =    \sum_{r=0}^{2K-1}
\sum_{s=0}^{2K-1} h_r h_s I_{2m+r,2n+s,+}
\label{eq:BAO}
\eeq
Because $I^k_{m+} = <x^k>_{0m}$ for $m\geq 0$, Eq. (\ref{eq:BAO}) becomes
a linear system for the unknown partial moments in terms of the full
moments.  These equations can be solved for the non-trivial $I^k_{m+}$.
The $ I^k_{m-}$ are obtained using 
$I^k_{m-}= I^k_{m} -  I^k_{m+}$.   Useful examples can be found in 
\cite{Shann1}\cite{Shann2} and \cite{web}.

For $I_{mn+}$ we have for $m \geq 0$ or $n \geq 0$:  
\beq
I_{mn+} = \delta_{mn} .
\label{eq:BAQ}
\eeq
from the orthonormality of the $\phi_{0l}(x)$.
This leads to a small linear system that relates the unknown integrals 
$I_{mn+}$ to the known $I_{mn}=\delta_{mn}$.  The $I_{mn-}$ can 
again be obtained by subtraction.

These linear systems can be solved for the non-trivial $I_{mn+}$.  The
corresponding partial moments on other scales can be obtained from the
$I_{mn+}$ using the unitarity of the operator $\hat{D}$ on the half
intervals.  Partial moments with different endpoints can be obtained
by translation and subtraction.  These integrals are used to treat
endpoint quadratures after the $K$-matrix equation is transformed to a
finite interval.

\noindent{\bf E. Singularity:}

The scaling equation can also be used to solve for the 
singular integrals of the form
\beq
I^{\pm}_k :=\int {\phi (x-k) \over x\pm i0^+} dx .
\label{eq:BAS}
\eeq
Application of the scaling equation (\ref{eq:BC}) leads 
to the linear relations:
\beq
I^{\pm}_k :=\sqrt{2} \sum_{l=0}^{2K-1} h_l I^{\pm}_{2k-l}. 
\label{eq:BAT}
\eeq
Using this equation for different values of $k$ leads to a set of linear 
equations that relate the $I^{\pm}_{k}$ for values of $k$ where 
the scaling basis function 
has support containing the singularity to the $I^{\pm}_{k}$ for values 
of $k$ where the scaling basis function has support far away from 
the singularity.   One more equation relating theses quantities 
is needed to pick up the contribution from singularity.  Specifically 
we use 
\beq
1 = \sum_n \phi_n (x), 
\eeq
which follows from Eq. (\ref{eq:BV}) to get:
\beq
\mp i \pi = \int_{-a}^a {dx \over x \pm i 0^+} = 
\sum_n \int_{-a}^a \phi_{n} (x) {dx \over x \pm i 0^+} =
\sum_n I^{\pm}_{n:a} ,
\label{eq:BAU}
\eeq
where
\beq
I^{\pm}_{n:a} = \int_{-a}^a {\phi (x-n) \over x\pm i0^+} dx = 
I^{\pm}_{n} 
\label{eq:BAV}
\eeq
except when the points $\pm a$ are in the support of $\phi_n(x) $.
The values $a$ is chosen to be far from the singularity.
For the principal value integral, which is used in this paper, we replace
$\mp i \pi$ on the left of Eq. (\ref{eq:BAU}) by zero. 

The input needed to solve this system are the values of 
$I^{\pm}_{n:a}$ and  $I^{\pm}_{n}$ far from the singularity.
For $\vert n \vert $ sufficiently large we can calculate 
these quantities by expanding them in terms of the moments
\beq
I^{\pm}_{n} = \int {\phi (x-n) \over x\pm i0^+}dx =
{1 \over n} \int {\phi (x) \over 1+ x/n }dx =
{1 \over n} \sum_{k=0}^{\infty} \left ({-1 \over n} \right )^k <x^k>_{00}
\label{eq:BAW}
\eeq
and the partial moments
\beq
I^{\pm}_{n:a} = \int_{-a}^a {\phi (x-n) \over x\pm i0^+}dx =
{1 \over n} \int_{-a}^a {\phi (x) \over 1+ x/n } dx=
{1 \over n} \sum_{k=0}^{\infty} \left ({-1 \over n} \right )^k <x^k>_{00:a}.
\label{eq:BAX}
\eeq
These integrals can be accurately approximated for sufficiently 
large $\vert n \vert$  
by truncating the series, (\ref{eq:BAW}) and (\ref{eq:BAX}),
for 
sufficiently large $\vert n\vert$.

Examples of the integrals that overlap the singularity are given in
Tables 3 and 4 for the $K=2$ and $K=3$ Daubechies wavelets.  The
endpoint integrals, where the singularity is located at the boundary
of the support, are not singular because of the continuity of $\phi
(x)$ for $K>1$.  These integrals, and integrals over scaling basis
functions close to the singularity are computed using the linear
equations (\ref{eq:BAT}) and (\ref{eq:BAU}). 
Singular integrals corresponding to different scales are
related by the identity
\beq
\int {\phi_{j,k} (x) dx\over x\pm i0^+} = 2^{j/2} I^{\pm}_{k} .
\eeq
The principal-value integrals are given by the real parts of $I^{\pm}_k$.
 
\begin{table} 
{\bf Table 3 - Integrals over singularity: K=2 $I^{\pm}_k$ } \\[1.0ex]
\begin{tabular}{|l|l|l|}
\hline
$I^{\pm}_{-1}$ & -2.779949550280  & $\mp i$  4.291495373146 \\
$I^{\pm}_{-2}$ & -0.269952669589  & $\pm i$  1.149902719556 \\
\hline					 
\end{tabular}					 
\end{table}

\begin{table} 
{\bf Table 4 - Integrals over singularity: K=3 $I^{\pm}_k$ } \\[1.0ex]
\begin{tabular}{|l|l|l|}
\hline
$I^{\pm}_{-1}$ &-0.1717835441734 & $\mp i$ 4.041140804162 \\
$I^{\pm}_{-2}$ &-1.7516314066967 & $\pm i$ 1.212142562305 \\
$I^{\pm}_{-3}$ &-0.3025942645356 & $\mp i$ 0.299291822651 \\
$I^{\pm}_{-4}$ &-0.3076858066180 & $\mp i$ 0.013302589081 \\
\hline					 
\end{tabular}					 
\end{table}

\section{Test Problem}

The power of the wavelet method is demonstrated by calculating the
$K$-matrix for an s-wave Malfliet-Tjon V potential.  The 
momentum-space integral equation for the two-body $K$-matrix is
\begin{equation} 
K_0(p_1,p_2,p_0) = v_0 (p_1, p_2) - m\, \mbox{P.V.}\int_0^{\infty} 
{v_0 (p_1, p')p^{\prime 2}  \over p^{\prime 2} -p_0^2}    
K_0(p',p_2,p_0) dp'
\label{eq:AA}
\end{equation} 
where $m$ is the nucleon mass and P.V. indicates the principal 
value.  The interaction potential is a sum of two Yukawa interactions
\begin{equation}
v_0 (p_1, p_2) :=
\sum_{i=1}^2 {\lambda_i \over 2 \pi p_1 p_2} \ln \left (
{\mu_i^2 + p_1^2 +p_2^2 + 2 p_1 p_2 \over 
\mu_i^2 + p_1^2 +p_2^2 - 2 p_1 p_2} \right )
\label{eq:AB}
\end{equation}
with strength and range parameters \cite{Payne} given in Table 2. 

\begin{table} 
{\bf  Table 2: Potential Parameters} \\[1.0ex]
$$\mbox{
\begin{tabular}{|c|c|c|c|c|}
\hline
$1/m$ & $\lambda_1$ & $\mu_1$ & $\lambda_2$ & $\mu_2$ \\
\hline
41.47 MeV fm$^2$  & -570.316 MeV fm & 1.55 fm$^{-1}$ & 1438.4812 MeV fm & 3.11 
fm$^{-1}$\\
\hline
\end{tabular}
}$$
\end{table}
Test calculations are done for the half on-shell $K$-matrix, $K_0 
(p_1,p_2,p_0)$, with $p_2=p_0$, and $p^2_0/m=10$ and $80$ MeV. 

\section{Wavelet Techniques for Singular Integral Equations}

We solve the integral equation (\ref{eq:AA}) by transforming 
the half-interval $[0,\infty)$ to a finite interval $[-a,b]$ with the
singularity $p_0$ transformed to the origin.

For fixed values of $p_2$ and $p_0$ define
\beq 
f(p_1):=  K( p_1,p_2,p_0)
\label{eq:CA}
\eeq
\beq
g (p_1) := v_0( p_1,p_2) .
\label{eq:CB}
\eeq
The fixed variables $p_2$ and $p_0$, which are just parameters in the
integral equation, are suppressed in this notation.

If we also define the non-singular part of the kernel 
\beq
L (p_1,p_2) := m  {v_0 (p_1, p_2) p_2^2 \over p_2 + p_0},
\label{eq:CC}
\eeq
the integral equation for the  $K$-matrix has the form
\beq
f(p_1) = g (p_1) -  \mbox{P.V.}\int_0^{\infty} 
{L(p_1,p')\over p'-p_0}  f(p') dp'.
\label{eq:CD}
\eeq

To transform $[0,\infty)$ to the interval $[-a,b]$ we use the 
following map, which also maps the singularity at $p_0$ to zero:
\beq
p= p(u) := p_0{b \over a} {a+u \over b-u} \qquad u = u(p):= 
{ab(p-p_0) \over a p+p_0 b}
\label{eq:CE}
\eeq
\beq
dp = p_0 {b \over a} { (b+a) \over (b-u)^2}du .
\label{eq:CF}
\eeq
\beq
{1 \over p-p_0} = {a(b -u) \over (a+b)p_0}{1 \over u}. 
\label{eq:CI}
\eeq
Let
\beq
\tilde{g}(u) := g \left ( p_0{b \over a} {a+u \over b-u} \right ),
\label{eq:CH}
\eeq
\beq
\tilde{f}(u) := f \left ( p_0{b \over a} {a+u \over b-u} \right )
\label{eq:CHX}
\eeq
and
\beq
\tilde{L}(u,v) = L \left (p_0{b \over a} {a+u \over b-u}, 
p_0{b \over a} {a+v \over b-v},p_0 \right ) 
{b \over (b-v)} ,
\label{eq:CJ}
\eeq
where the factor $b/(b-v)$ is the product of the Jacobian and
the residue of the transformed singularity   
\beq
{b \over b-v}=
p_0 {b \over a} { (b+a) \over (b-v)^2}
{a(b -v) \over (a+b)p_0}.
\label{eq:CK}
\eeq

The integral equation in the $u,v$ variables has the form
\beq
\tilde{f}(u) = \tilde{g}(u) + \int_{-a}^b 
{\tilde{L}(u,v) \over v}  \tilde{f}(v)  dv .
\label{eq:CL}
\eeq
To solve this we represent $\tilde{f}(u)$ using the scaling basis
on a sufficiently fine scale $j$:
\beq
\tilde{f} (u) \approx \sum_n \tilde{f}_n \phi_{j,n} (u).  
\label{eq:CM}
\eeq
Inserting the expansion (\ref{eq:CM}) in the integral  
equation (\ref{eq:CL}) gives 

\beq
\sum_n \phi_{j,n} (u) \tilde{f}_n = \tilde{g}(u) + \sum_n  \int_{-a}^b 
{\tilde{L}(u,v) \over v}  \phi_{j,n} (u) dv \, \tilde{f}_n .
\label{eq:CN}
\eeq
To get a linear system for the coefficients $f_n$ we multiply the 
above equation by $\phi_{j,m}(u)$ and integrate from $-a$ to $b$: 
\beq
\sum_n  N_{mn} \tilde{f}_n  = \tilde{g}_{m} + \sum_n  \int_{-a}^b 
\phi_{j,m}(v) {\tilde{L} (u,v) \over v}  \phi_{j,n} (u)\, du \,dv \,
\tilde{f}_n ,
\label{eq:CO}
\eeq
where 
\beq
N_{mn} := \int_{-a}^b \phi_{j,m} (u) \phi_{j,n} (u) du. 
\label{eq:CP}
\eeq
The matrix $N_{mn}$ is $\delta_{mn}$ except when the support of 
$\phi_{j,m} (u)$ and $\phi_{j,n} (u)$
contains $-a$ or $b$. This is expressed in the form 
\beq
N_{mn}= \delta_{mn} + \Delta_{mn} .
\label{eq:CQ}
\eeq
The integrals $N_{mn}$ can be computed exactly using
Eq. (\ref{eq:BAJ}-\ref{eq:BAQ}).  The integrals, $N_{mn}$, the
singular integrals, $I^{\pm}_m$, as well as the moments, partial
moments, and quadrature points can be calculated once and stored
for later use.

The vector $\tilde{g}_{m}$ is given by
\beq
\tilde{g}_{m}= \int_{-a}^b \tilde{g}(u)\phi_{j,m} (u) du ,
\label{eq:CR}
\eeq
which is computed using the one-point quadrature when $-a$ or $b$ is not
in the support of $\phi_{j,m}(u)$.  In the case that $-a$ or $b$ is in
the support of $\phi_{j,m}$ we use the partial moments
(\ref{eq:BAH}) to construct endpoint quadrature rules (\ref{eq:BADX}).
We use $K+1$ transformed Gauss-Legendre points.  The weights are
determined by solving a linear equation determined by the requirement
that the quadrature rule reproduces the lowest-order partial moments
up to order $K$.
  
The matrix elements of the kernel are evaluated using
\[
\tilde{L}_{m,n}:=
\int_{-a}^b 
\phi_{j,m}(v) {\tilde{L}(v,u) \over u}  \phi_{j,n} (u) \,du\, dv  
\]
\[
=\int_{-a}^b 
\phi_{j,m}(v) {\tilde{L}(v,u)-\tilde{L}(v,0) \over u}  \phi_{j,n} (u)\, du\, dv
\]
\beq
+ \int_{-a}^b 
\phi_{j,m}(v) \tilde{L}(v,0) dv\int_{-a}^b {\phi_{j,n} (u)  \over u}  du  .
\label{eq:CS}
\eeq

For most basis functions the $v$ integral is done using the one-point
quadrature rule.  This is also the case in Eq. (\ref{eq:CR}) for the
``$u$'' integral and for the integral over the subtracted term in
Eq. (\ref{eq:CS}).  The quadratures used for the endpoints of
$\hat{g}_m$ are used to integrate scaling functions with support on
the endpoints of the integral.  The singular ``$u$'' integral on the
last line of Eq. (\ref{eq:CS}) is done using the method discussed in
the previous section.

The resulting equation has the form
\beq
\tilde{f}_m  = \tilde{g}_m + 
\sum_n (\tilde{L}_{m,n} - \Delta_{m,n} ) \tilde{f}_n.
\label{eq:CT}
\eeq
The coefficients $\tilde{f}_m$ obtained by solving this matrix equation 
can be used to calculate a refined solution,  which is constructed by
substituting the series solution into the right-hand side of the 
integral equation.  If the integrals are evaluated following 
the same procedure used in the integral equations, the 
solution $\tilde{f}(u)$ can be obtained for 
any value of $u$ without having to calculate the basis functions
$\phi_{j,m}(u)$.   This is significant because
the evaluation of the basis functions is computationally intensive.
The refined solution, $f(u)$, can be used to evaluate the half 
on-shell $K$-matrix at any point $p$ by using the inverse of (\ref{eq:CHX}):
\beq
K(p,p_0,p_0) = f (p) = \tilde{f} 
\left ({ab(p-p_0) \over a p +p_0 b}  \right )
\label{eq:CG}
\eeq   
 
The calculation outlined above, where the transformed  
K-matrix equation is solved in the scaling function basis, has 
a full kernel matrix.  While the treatment of the singular integral is 
very stable and accurate,  the method does not lead to a reduction
in the size of the matrix equations when compared to calculations using 
splines or other numerical techniques.
In the next section we show that if Eq. (\ref{eq:CT}) is transformed to 
an equivalent equation in the wavelet basis, that the resulting 
kernel matrix is well approximated by a sparse matrix, which leads to a
considerable reduction in the scale of the numerical calculation.  
We also show that the wavelet transform can be implemented by an
efficient algorithm that uses a simple filter on multiple scales.

\section{Wavelet Transform} 

The computation done in the previous section leads to a linear system
with a large dense matrix.  The solution is a set of numerical
coefficients $\tilde{f}_l = c_{j,l}$ of the approximate solution in
the scaling function basis for a fine scale $j$.
As mentioned earlier, there are two equivalent representations for the
solution $\tilde{f} (u)$ given in Eq. (\ref{eq:BP}).
The wavelet transform for a problem with $2^N$ basis functions maps
coefficients of the basis functions on model space ${\cal V}_j$ to
coefficients of the basis functions on the equivalent space ${\cal
W}_{j+1}\oplus \cdots {\cal W}_{j+N}\oplus {\cal V}_{j+N}$.  

The form of the transformation is determined by applying the 
Eqs. (\ref{eq:BC}) and (\ref{eq:BK}) to map
\beq
{\cal V}_l \to {\cal V}_{l+1}
\eeq
and 
\beq
{\cal V}_l \to {\cal W}_{l+1}.
\eeq
This is done $N$ times, successively applying Eqs. (\ref{eq:BC}) 
and (\ref{eq:BK})
to ${\cal V}_j$, ${\cal V}_{j+1} \cdots {\cal V}_{j+N-1}$. 

The coefficients on the next (coarser) scale are obtained using 
\beq 
c_{j+1,m}= \sqrt{2} \sum_l h_l c_{j,m+l} 
\label{eq:DA}
\eeq
\beq 
d_{j+1,m}= \sqrt{2} \sum_l g_l c_{j,m+l} .
\label{eq:DB}
\eeq 
Because only a finite number of coefficients are considered, there is 
some ambiguity in how to treat the coefficients that appear on the 
right-hand side of these equations, which are out of range.   The simplest
method to treat these terms is to use a periodic wrap-around condition:
\beq
c_{j+k,l+ 2^{N-k}} \to c_{j+k,l}\,,
\eeq
which is applied at each level, $k$. Since the result is an orthogonal
transformation of the original basis, this choice does not affect the 
solution.   It will have a small edge effect on the structure of 
the resulting sparse matrix. 

The structure of the successive transformations is illustrated for the
case $j=-3$ and $N=3$ by the following diagram. In this diagram
$c_{j,m}$ represents the coefficients of the basis functions in ${\cal
V}_j$ and $d_{j,m}$ represents the coefficients of the basis functions
in ${\cal W}_j$.  The successive transformations act on $c_{-3,l}$ as follows:
\beq
\left (
\begin{array}{c}
c_{-3,1} \\
c_{-3,2} \\
c_{-3,3} \\
c_{-3,4} \\
c_{-3,5} \\
c_{-3,6} \\
c_{-3,7} \\
c_{-3,8} 
\end{array}
\right )
\to 
\left (
\begin{array}{c}
c_{-2,1} \\
c_{-2,2} \\
c_{-2,3} \\
c_{-2,4} \\
d_{-2,1} \\
d_{-2,2} \\
d_{-2,3} \\
d_{-2,4} 
\end{array}
\right )
\to
\left (
\begin{array}{c}
c_{-1,1} \\
c_{-1,2} \\
d_{-1,1} \\
d_{-1,2} \\
d_{-2,1} \\
d_{-2,2} \\
d_{-2,3} \\
d_{-2,4} 
\end{array}
\right )
\to 
\left (
\begin{array}{c}
c_{-0,1} \\
d_{-0,1} \\
d_{-1,1} \\
d_{-1,2} \\
d_{-2,1} \\
d_{-2,2} \\
d_{-2,3} \\
d_{-2,4} 
\end{array}
\right )
\eeq
This transform can be implemented efficiently.  Since it is an
orthogonal transformation, if it is expressed in matrix form, the
inverse is just the transpose.  Both the wavelet transform and its
inverse can be implemented in $O(N)$ \cite{Resnik} steps.

The advantage of using the wavelet basis is that the coefficients $d_{j,k}$ in
Eq. (\ref{eq:BP}) will be small if $\int \tilde{f} (u) \psi_{j,k}(u)\, du
\approx 0$.  This will be true, by virtue of Eq. (\ref{eq:BT}), 
whenever $\tilde{f}(u)$ can be locally approximated by
a polynomial of degree less than $K$ on the support of
$\psi_{j,k}(u)$.  A practical rule of thumb states that the number of
non-zero matrix elements grows like $10 N \log (1 / \epsilon)$
\cite{numrec}, where $N$ is the size of one dimension of the matrix,
and $\epsilon$ is the absolute value of the ratio of size of the
smallest matrix element retained to the largest matrix element.
This behavior is consistent with our observations.

In Figure 3 a simple example is used to illustrate how structure
appears on different scales once the wavelet transform is applied.  In
this case a product of two Gaussian functions with displaced broad and
narrow peaks is expressed in a K=3 Daubechies scaling basis.  The
result is transformed with the wavelet transform and the strength of
the coefficients in the wavelet basis are studied.  The figure shows
the coefficients of the scaling functions on the coarse scale give a
low-resolution large-scale representation of the original function.
The coefficients of the wavelets on different scales builds
the fine structure.  Note the that scales on the vertical axes differ
in each frame.  It is also interesting to see how the wavelet
transform automatically finds the parts of the original function with
structure.

To fully use the wavelet transform, the number of basis functions
should be a power of two.  In this paper the choice of $-a$ and $b$ is
adjusted so that the number of basis functions is a power of two.  By
translating the interval $[-a,b]$ it is also possible to control the
number of basis functions with support on either side of the
singularity at zero.

In applications the wavelet transformation is applied to the driving
term and to the rows and columns of the kernel of the integral
equation.  This leads to an equivalent linear system with respect to 
the transformed basis.  In this new basis we expect the matrix of the 
kernel to be the sum of a sparse matrix and a matrix with small norm.
Ignoring the small normed matrix leads to an approximate linear system 
with a sparse kernel.

This approximate solution can be transformed back to the scaling
representation and substituted back into the integral equation.  
Evaluating the integrals using the one-point
quadrature and methods discussed previously, gives the refined solution,
constructed without computing the wavelets or scaling basis functions.

\section{Results} 

Initial calculations are performed using  K=2 and K=3 Daubechies
wavelets using $N$ scaling basis functions,  up to $N_{max}= 2^9=512$.
The finest resolution corresponds to $j:=J=-7$.  The values of $a$ and 
$b$ are given by $a=1$ and $b=-a+(N-2K+2)2^{J}$.  
The solutions are represented directly in terms of a series expansion
in the scaling function basis and using this expansion in the
integral equation to construct the refined solution.  The refined
solution has the advantages that the fractal structure of the basis
does not appear in the solution and the basis functions do not have to
be evaluated.  These calculations were checked against conventional
differential and integral equation programs, and the agreement was
excellent.

Calculations were done for the half on-shell K-matrix at 10 and 80
MeV.  Tables 5-8 show the values of the refined on-shell K-matrix elements
at 10 and 80 MeV for the Daubechies $K=2$ and $K=3$ scaling function
bases for different resolutions $j$ (approximation spaces ${\cal V}_j$).  
These tables
illustrate the convergence of a full calculation as the resolution
is increased.  In each successive calculation the
resolution is increased by a factor of 2.

The convergence for the $K=2$ and $K=3$ calculations are compared for
different values of $j=J$, which defines the smallest scale, in Tables
9-10.  It is apparent that both methods are accurate, but the $K=3$
Daubechies basis gives better results for the same number of basis
functions.

In this paper the refined solution with $N=2^9$ scaling basis functions
serves as the benchmark calculation that is used to test the
approximate wavelet calculations with sparse matrices.  In general,
the method leads to accurate and stable solutions.  This solution is
converged to the accuracy expressed in the tables and agrees with 
calculations using other methods.

Applying the wavelet transform to the integral equation and throwing
away all matrix elements of the kernel smaller than a given size leads
to approximate equations with sparse matrices.  Tables 11-13 compare
the solution of different sparse matrix approximations to the full 512
basis function calculation for the $K=2$ and $K=3$ Daubechies basis
functions at 10 and 80 MeV.  The first column in each of these tables
indicates that matrix elements with an absolute value smaller than
$\epsilon$ times the absolute value of the largest matrix elements are
set to zero.  The second column indicates the percentage of the $512
\time 512$ matrix elements that are non-zero in the sparse matrix
approximation.  The third column indicates the on-shell value of the
$K$-matrix obtained by applying the integral equation refinement to
the sparse matrix solution.  Column five shows the relative error
at the on-shell point and column six shows the mean square error of 
solution to the
integral equation over the whole domain defined by:
\beq
\left \vert {\tilde{f}(0)-\tilde{f}_\epsilon (0) \over \tilde{f} (0)}
\right \vert
\label{eq:EA}
\eeq
and
\beq
{ \int (\tilde{f}(u)-\tilde{f}_\epsilon (u))^2 du )^{1/2} \over
( \int  \tilde{f}^2(u)  du )^{1/2} }.
\label{eq:EA}
\eeq
The power of the wavelet method is apparent in the $K=3$ Daubechies
basis.  Tables 12 and 14 show, for the case $\epsilon = 10^{-6}$, 
that about
96\% of the matrix elements are eliminated relative to the full
calculations.  The on-shell error is a few parts in $10^5$
for both energies.  Even more impressive is the fact that the 
mean-square error for the half-shell $K$-matrix is also about the same
order of magnitude.

Figure 4 shows the transformed half on-shell K-matrix, $\tilde{f}(u)$
at 80 MeV using the sparse matrix approximation with $\epsilon =
10^{-6}$.  This figure is a plot of the series solution, which is
indistinguishable from the refined solution.  The smoothness of the
series solution is apparent from this figure.  While in Figure 5 
we show the half on-shell K-matrix, $f(p_1)$, at 80 MeV using the sparse matrix
approximation with $\epsilon = 10^{-6}$.  This plot shows the refined
solution, which is obtained by substituting the series solution back
in the integral equation.  As in the transformed case, this is
indistinguishable from the series solution.  Figure 6 provides a
graphic illustration of how the non-zero terms of the sparse kernel
are distributed for the case $\epsilon = 10^{-6}$.  What is relevant
is that the two dimensional plot is dominated by a number of
one-dimensional structures.  The vertical bands in this matrix
correspond to basis functions with support near the singularity, where
we expect significant structure.  The different diagonal bands 
arise because local structures in the kernel can couple wavelets on 
several scales.
     
While it is possible to calculate perturbative corrections to
approximate solutions, it make more sense calculate the matrix
equation in the scaling basis at a finer scale, and use the wavelet
transform to get a better approximation.  This is because the
matrix with small norm, that is needed to compute perturbative 
corrections, is not
sparse.

The tables and calculations indicate that wavelet methods
provide a potentially powerful technique for reducing the size of
momentum-space scattering integral equations.  In this paper we did
this by transforming a dense kernel using the wavelet transform.
These operations can be done very efficiently, resulting in a matrix 
which is the sum of a sparse matrix and a matrix with small norm.

Solving the linear system using only the sparse matrix leads to 
accurate solutions.  In our application a decrease in the size of the 
matrix by a factor of 20 led to an error of a few parts in $10^5$
for the Daubechies $K=3$ wavelets.

The speed of our computations is helped by using the one-point
quadrature, which leads to accurate answers when used with 
the Daubechies $K=2$ or 3 wavelets.  Another nice feature of the 
wavelet method is that it automatically puts basis functions at places
where the kernel has structure, as compared to splines, where the 
placement of the knots must be done by hand.
 
\section{Summary}

In this paper we adapted numerical methods based on wavelets to calculate
the K-matrix for a Malfliet-Tjon V potential in momentum space.
The motivation for this work was to determine if numerical methods based on
wavelets could be used to reduce the size of the matrices that
occur in numerical implementations of the momentum-space integral 
scattering equations.

There are three reasons for investigating this problem:
\begin{itemize}

\item[1.] Relativistic formulations of the few-body problem are naturally
formulated in momentum space.  This is because the momentum operators
are infinitesimal generators of the Poincar\'e group.  Operators,
like Wigner rotations, have simple forms in momentum space. 

\item[2.] Boundary conditions 
and scattering asymptotic conditions can be cleanly 
formulated in momentum-space compact-kernel integral equations.

\item[3.] The kernel matrix that appears in most numerical implementations of 
momentum-space scattering equations is full.  The size of the 
kernel matrix is also large for few-body calculations with 
realistic interactions.
\end{itemize} 

We found that the matrix representation of the
momentum-space scattering integral equations 
in the wavelet basis leads to a linear system with
a sparse approximate kernel matrix.  This leads to a significant
reduction in the size of the dynamical calculation.

Our applications used the Daubechies $K=2$ or
$K=3$ wavelets.  For each value of $K$ there were two bases, the
scaling function basis and the wavelet basis, which were related by a
fast orthogonal transform, called the wavelet transform.  The $K=2$
bases can locally represent lines of arbitrary slope on subintervals
corresponding to the finest scale.  The $K=3$ bases can locally
represent quadratic curves on subintervals corresponding to the finest
scale.  Our observations indicate that we obtain better accuracy with
the same number of basis functions using the $K=3$ basis than with the
$K=2$ basis without a significant increase in computational
complexity.  While it is possible to use a basis with a higher $K$
value, in order to take advantage of the additional capabilities of
the basis functions, the one-point quadrature would have to be
replaced by a multi-point quadrature.  The additional benefit is not
worth the complexity and increased computational effort 
required by the additional quadrature points.

One potential difficulty with using wavelet methods in numerical
calculations is the self-similar structure of the basis functions.
Functions that do not become smooth on a sufficiently small scale are
not amenable to conventional numerical methods.  Initially, we found
that this property caused numerical challenges. To overcome these
problems we utilized the scaling equation to {\it exactly} calculate
moments of the scaling basis functions or a wavelets times a
polynomial.  We found that by using these moments and the scaling
equation we were able to exactly compute overlap integrals, to
accurately evaluate integrals over the scattering singularity, and to
exactly compute endpoint overlap integrals.  It is only after using
the scaling equation to do all of the required quadratures that we
were able to obtain stable and accurate answers.  While many of the
techniques discussed in this paper are standard wavelet procedures,
the method for treating the scattering singularity seems to be new to
this paper. 
 
As a basis for solving integral equations, the scaling function basis
has many of the same advantages as a spline basis.  The most important
common features are the compact support and the ability to exactly
represent polynomials on small scales.  This means the they can be
used to efficiently approximate local structures.  The scaling
function basis has the additional feature that the functions are
orthonormal.  In addition, for the $K=2,3$ Daubechies wavelets, the
one-point quadrature can be used to exactly compute the integral of the
product of a degree two polynomial and the scaling basis function on
the support of the scaling basis function.
By employing the one-point quadrature rule, the matrix formulation of 
the integral equation can be computed very quickly.   While the wavelets 
and scaling basis functions are computationally intensive to compute,  
the one-point quadrature can be used to replace the evaluations of 
these functions if the series solutions are substituted back in the 
integral equation.  This method, leads to our refined solution.  The only 
input that is needed are the scaling coefficients, $h_l$, which are
known analytically for the $K=2$ and $K=3$ wavelets.

In working on this problem, we were able to find a number of previous
applications of wavelet methods to integral equations; but were unable
find applications that treated integral equations with the singular
kernels of momentum space scattering equations.  We found that the
required integrals could be accurately computed using the scaling
equation.  The method has proved to be very stable.  It has the
advantage that it reduced the calculation of these integrals to
solving a small system of linear equations,
the results of which can be stored for later calculations.

The real advantage of the wavelet method over the conventional methods
occurs after the wavelet transform is applied to the K-matrix equation
in the scaling function basis.  The kernel matrix in the wavelet 
basis can be expressed as the sum of sparse matrix and a
matrix of small norm.  Ignoring the small matrix leads to an
approximate solution.  The case featured in Figure 6, used Daubechies
$K=3$ wavelets.  The sparse matrix approximation used in Figure 6 led
to a 96\% reduction in the number of non-zero matrix elements.  This
was done with a dimensionless mean square error of a few parts in
$10^5$

The beauty of the wavelet transform is that it is fast, and it
automatically finds the basis functions needed to represent the small
scale structure of the solution.
 
In this paper we have exploited the scaling equation to formulate a
powerful method for solving momentum space scattering integral
equations.  While we limited our application to what was necessary to
solve the scattering integral equations, it is apparent that there are
many thing that can be computed exactly or to a very high accuracy
with the scaling equation.  For example, it is possible to exactly
integrate products of several scaling functions and powers of the
variable.  This suggests that these methods could be very useful for
solving a large class of physics problems.

Our conclusion is that wavelet bases provide a powerful method for
solving momentum-space scattering integral equations.  The required
matrix elements are easy to compute, and the wavelet transform can be
implemented very efficiently, leading to an approximate linear system
with a sparse matrix.
 
\begin{table} 
{\bf Table 5 - Dependence on number of basis functions: K=2, E= 10 MeV} \\[1.0ex]
\begin{tabular}{|l|l|l|l|}
\hline
-J   &	N  & series on-shell&  refined on-shell \\ 
\hline					    		      
3   &	32    &  -124.681226   	&  -124.401416       \\   
4   &	64    &  -124.924473    &  -124.853374       \\  
5   &	128   &  -124.984907    &  -124.967026       \\   
6   &	256   &  -124.999853 	&  -124.995372       \\   
7   &	512   &  -125.003567 	&  -125.002445       \\    
\hline					 
\end{tabular}					 
\end{table}

\begin{table} 
{\bf Table 6 - Dependence on number of basis functions: K=3, E=10 MeV} \\[1.0ex]
\begin{tabular}{|l|l|l|l|}
\hline
-J  &	N      &  series on-shell&  refined on-shell \\
\hline								 
3  &	32	&  -125.051451   &   -125.034060 	\\
4  &	64	&  -125.007967	 &   -125.006049 	\\
5  &	128	&  -125.005171	 &   -125.004948 	\\
6  &	256	&  -125.004847	 &   -125.004820 	\\
7  &	512	&  -125.004806	 &   -125.004803 	\\
\hline
\end{tabular}
\end{table}

\begin{table} 
{\bf Table 7 - Dependence on number of basis functions: K=2, E=80 MeV} \\[1.0ex]
\begin{tabular}{|l|l|l|l|}
\hline
-J  &	N   & series on-shell & refined on-shell \\
\hline					      		      
3  &	32  &   -6.53375948	   &    -6.38393342		\\
4  &	64  &   -6.45483277	   &    -6.41711946		\\
5  &	128 &   -6.43490787	   &    -6.42555390		\\
6  &	256 &   -6.42998750	   &    -6.42766546		\\
7  &	512 &   -6.42877076	   &    -6.42819277		\\
\hline
\end{tabular}
\end{table}

\begin{table} 
{\bf Table 8 - Dependence on number of basis functions: K=3, E=80 MeV} \\[1.0ex]
\begin{tabular}{|l|l|l|l|}
\hline
-J  & N & series on-shell & refined on shell  \\
\hline					      		      
3   &	32  &   -6.44161445	 &  -6.43154124	 \\
4   &	64  &   -6.42926712	 &  -6.42868443	 \\
5   &	128 &   -6.42842366	 &  -6.42840177	 \\
6   &	256 &   -6.42837147	 &  -6.42837210	 \\
7   &	512 &   -6.42836848	 &  -6.42836877	 \\
\hline
\end{tabular}
\end{table}

\begin{table} 
{\bf Table 9 - Refined on-shell values, E=10 MeV } \\[1.0ex]
\begin{tabular}{|l|l|l|}
\hline
N & K=2 on-shell & K=3 on-shell  \\
\hline					      		      
32   &  -124.401416  &  -125.034060  \\
64   &  -124.853374  &  -125.006049  \\
128  &  -124.967026  &  -125.004948  \\
256  &  -124.995372  &  -125.004820  \\
512  &  -125.002445  &  -125.004803  \\
\hline
\end{tabular}
\end{table}

\begin{table} 
{\bf Table 10 - Refined on-shell values, E=80 MeV } \\[1.0ex]
\begin{tabular}{|l|l|l|}
\hline
N & K=2 on-shell & K=3 on-shell  \\
\hline					      		      
32   &  -6.38393342  &  -6.43154124 \\
64   &  -6.41711946  &  -6.42868443 \\
128  &  -6.42555390  &  -6.42840177 \\
256  &  -6.42766546  &  -6.42837210 \\
512  &  -6.42819277  &  -6.42836877 \\
\hline
\end{tabular}
\end{table}

\begin{table} 
{\bf Table 11 - Sparse Matrix Convergence: K=2, E=10 MeV, J=-7 } \\[1.0ex]
\begin{tabular}{|l|l|l|l|l|}
\hline
$\epsilon$ & percent & on-shell value &	on-shell error & mean-square error \\
\hline					    		        
0     &	100           &	-125.00245 	  &	0         &  	0	          \\
$ 10^{-9}$ &	34.24 &	-125.00244	  &	6.94$\times 10^{-9}$  &	8.76$\times 10^{-9}$	  \\
$10^{-8}$ &	20.8  &	-125.00244	  &	3.82$\times 10^{-8}$  &	7.81$\times 10^{-8}$	  \\
$10^{-7}$ &	12.16 &	-125.00440	  &	1.57$\times 10^{-5}$  &	1.63$\times 10^{-5}$      \\
$10^{-6}$ &	6     &	-124.99271	  &	7.78$\times 10^{-5}$  &	9.53$\times 10^{-5}$      \\
$10^{-5}$ &	2.61  &	-124.96069	  &	.000334	  &      .000435          \\
$10^{-4}$ &	1.16  &	-124.47119 	  &	.00425	  &      .00457           \\
$10^{-3}$ &	.55   &	-120.42759	  &	.0366	  &      .0415            \\
$10^{-2}$ &	.31   &	-116.14345	  &	.0709	  &      .154             \\
\hline
\end{tabular}
\end{table}

\begin{table} 
{\bf Table 12 - Sparse Matrix Convergence: K=3, E=10 MeV, J=-7 } \\[1.0ex]
\begin{tabular}{|l|l|l|l|l|}
\hline
$\epsilon$ & percent & on-shell value & on-shell error & mean-square error \\
\hline					    		        
0     &	100    &       	-125.00480  &	0	  &     0	          \\
$10^{-9}$ &	17.78 &	-125.00480  &	1.05$\times 10^{-8}$  &	2.56$\times 10^{-8}$	  \\
$10^{-8}$ &	11.38 &	-125.00480  &	5.14$\times 10^{-8}$  &	2.44$\times 10^{-7}$	  \\
$10^{-7}$ &	6.6   &	-125.00475  &	4.49$\times 10^{-7}$  &	1.88$\times 10^{-6}$	  \\
$10^{-6}$ &	3.76  &	-125.00269  &	1.69$\times 10^{-5}$  &	2.08$\times 10^{-5}$	  \\
$10^{-5}$ &	2.14  &	-124.99030  &	.000116   &	.000228		  \\
$10^{-4}$ &	1.24  &	-124.85112  &	.00123    &	.00217		  \\
$10^{-3}$  &	.72   &	-123.82508  &	.00944    &	.0117		  \\
$10^{-2}$   &	.38   &	-125.25766  &	.00202    &	.128		  \\
\hline
\end{tabular}
\end{table}

\begin{table} 
{\bf Table 13 - Sparse Matrix Convergence: K=2, E= 80, MeV J=-7 } \\[1.0ex]
\begin{tabular}{|l|l|l|l|l|}
\hline
$\epsilon$ & percent & on-shell value & on-shell error & mean-square error \\
\hline					    		        
0         &	100   & -6.4281928  &    0                     &    	0      	  \\
$10^{-9}$ &	38.25 &	-6.4281928  &	4.16$\times 10^{-11}$   &	6.30$\times 10^{-9}$       \\
$10^{-8}$ &	23.59 &	-6.4281927  &	1.54$\times 10^{-8}$    &	1.13$\times 10^{-7}$	  \\
$10^{-7}$ &	13.84 &	-6.4281936  &	1.30$\times 10^{-7}$	&       2.11$\times 10^{-6}$	  \\
$10^{-6}$ &	6.89  &	-6.4282684  &	1.18$\times 10^{-5}$    &	3.87$\times 10^{-5}$	  \\
$10^{-5}$ &	2.91  &	-6.4286669  &	7.38$\times 10^{-5}$    &	.000211		  \\
$10^{-4}$ &	1.18  &	-6.4286269  &	6.75$\times 10^{-5}$    &	.00166		  \\
$10^{-3}$ &	.55   &	-6.4083948  &	.00308	                &       .0113		  \\
$10^{-2}$  &	.3    &	-6.3206920  &	.0167                   &	.101		  \\
\hline	       
\end{tabular}
\end{table}

\begin{table} 
{\bf Table 14 - Sparse Matrix Convergence: K=3, E=80, MeV J=-7 } \\[1.0ex]
\begin{tabular}{|l|l|l|l|l|}
\hline
$\epsilon$ & percent & on-shell value & on-shell error & mean-square error \\
\hline					    		        
0     &	100   &	       	-6.4283688   &	0	 &       0		  \\
$10^{-9}$&	19.99 &	-6.4283688   &	1.52$\times 10^{-10}$  & 1.20$\times 10^{-8}$		  \\
$10^{-8}$ &	12.94 &	-6.4283690   &	3.44$\times 10^{-8}$ &	2.06$\times 10^{-7}$	  \\
$10^{-7}$ &	7.42  &	-6.4283703   &	2.33$\times 10^{-7}$ &	1.87$\times 10^{-6}$	  \\
$10^{-6}$ &	4.08  &	-6.4283333   &	5.51$\times 10^{-6}$ &	4.38$\times 10^{-5}$	  \\
$10^{-5}$ &	2.22  &	-6.4278663   &	7.82$\times 10^{-5}$ &	.000994		  \\
$10^{-4}$ &	1.21  &	-6.4244211   &	.000614  &	.00845		  \\
$10^{-3}$  &	.67   &	-6.4154328   &	.00201   &	.0229		  \\
$10^{-2}$  &	.34   &	-6.2935398   &	.021    &	.102		  \\
\hline
\end{tabular}
\end{table}

\vfill\eject

\noindent{\bf Figure 1: Daubechies $K=2$ Scaling function and mother wavelet 
function}

\includegraphics[5cm,7cm][17.5cm,21cm]{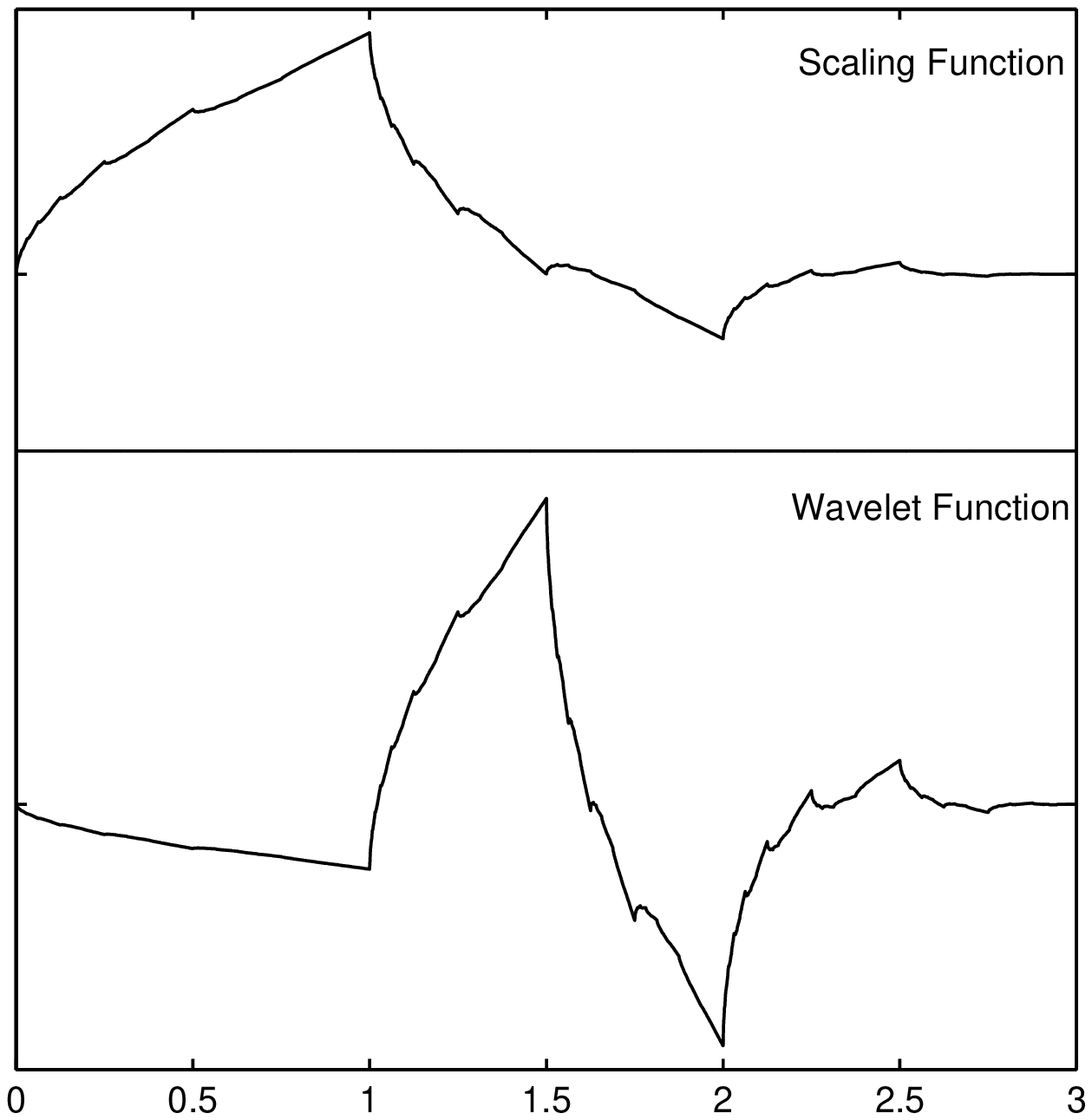}

\vfill\eject

\noindent{\bf Figure 2: Daubechies $K=3$ Scaling function and mother wavelet 
function}

\includegraphics[5cm,7cm][17.5cm,21cm]{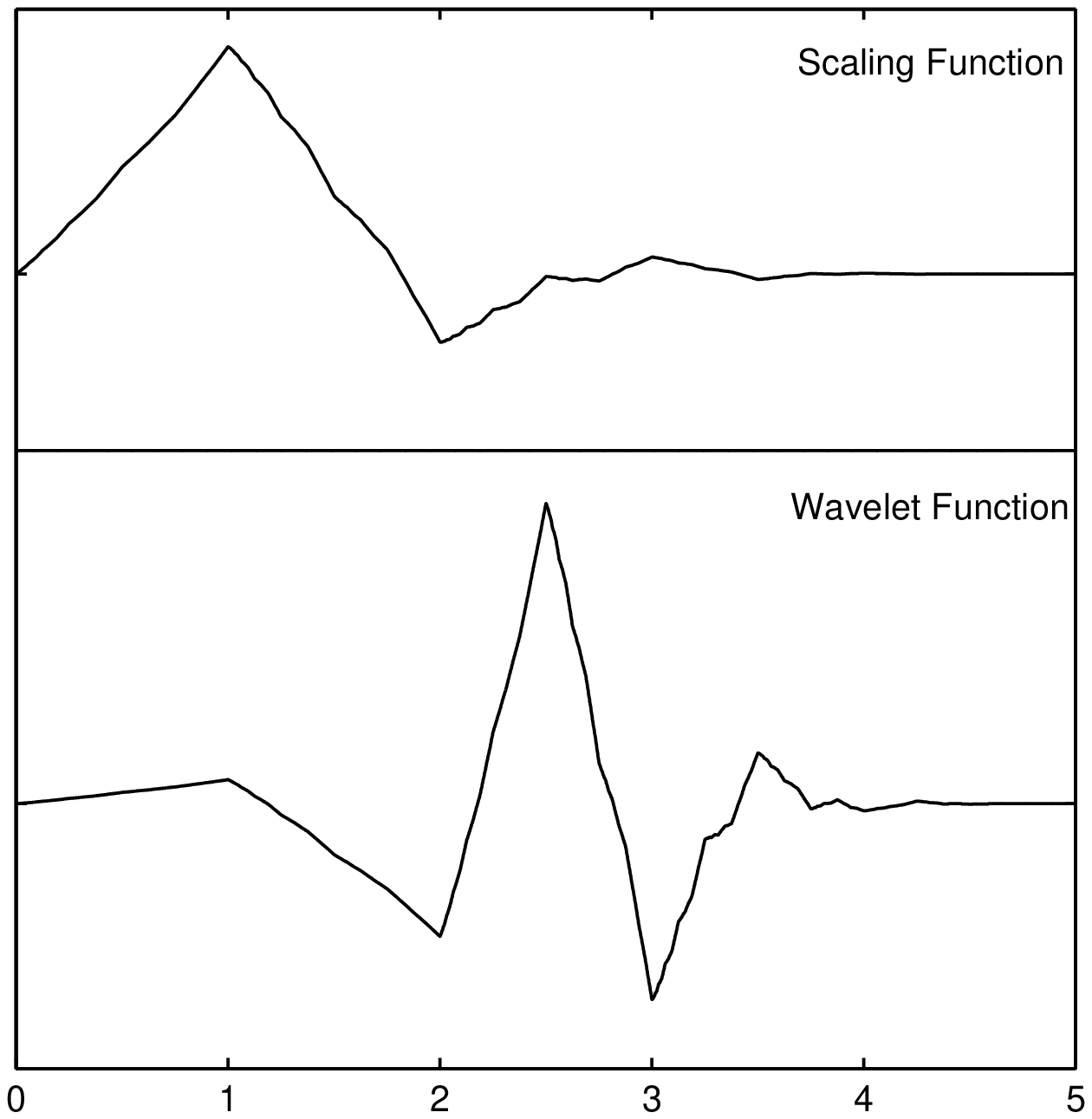}

\vfill\eject

\noindent{\bf Figure 3: Distribution of wavelet coefficients}

\includegraphics[5cm,7cm][17.5cm,21cm]{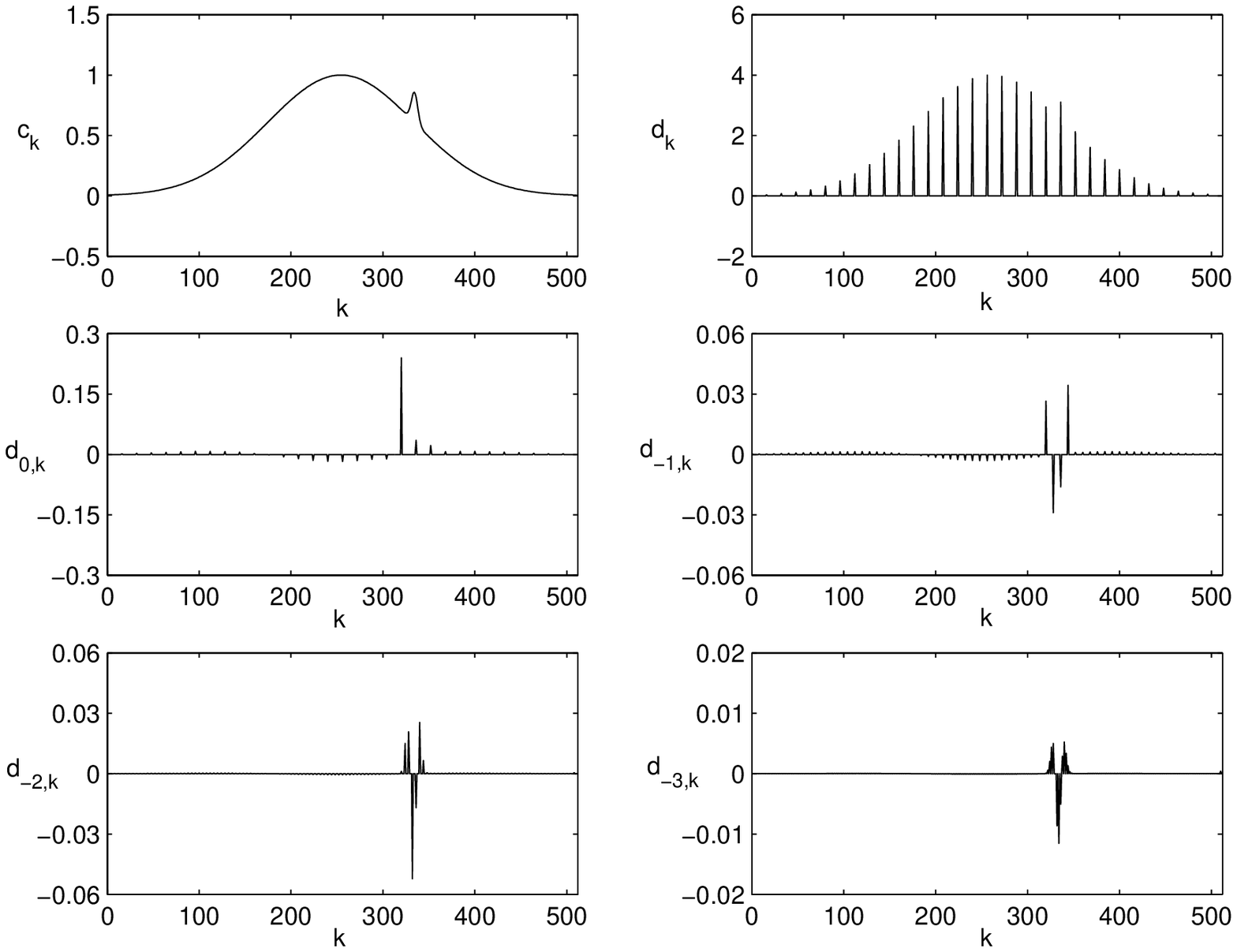}

\vfill\eject

\noindent{\bf Figure 4: Transformed K-Matrix - E = 80 MeV - expansion in wavelet basis}

\includegraphics[5cm,7cm][17.5cm,21cm]{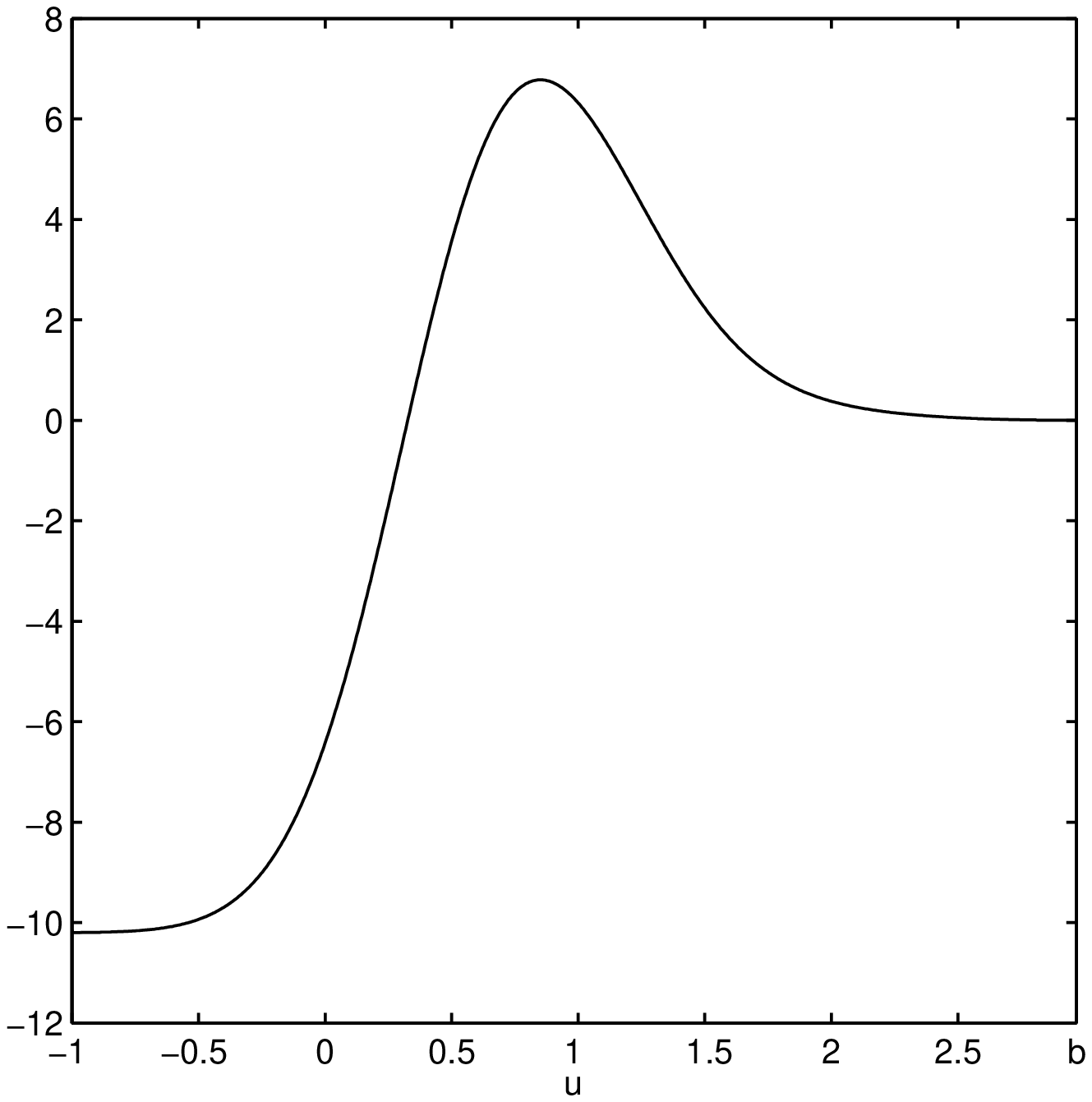}

\vfill\eject

\noindent{\bf Figure 5: K-Matrix - E = 80 MeV - refined solution}

\includegraphics[5cm,7cm][17.5cm,21cm]{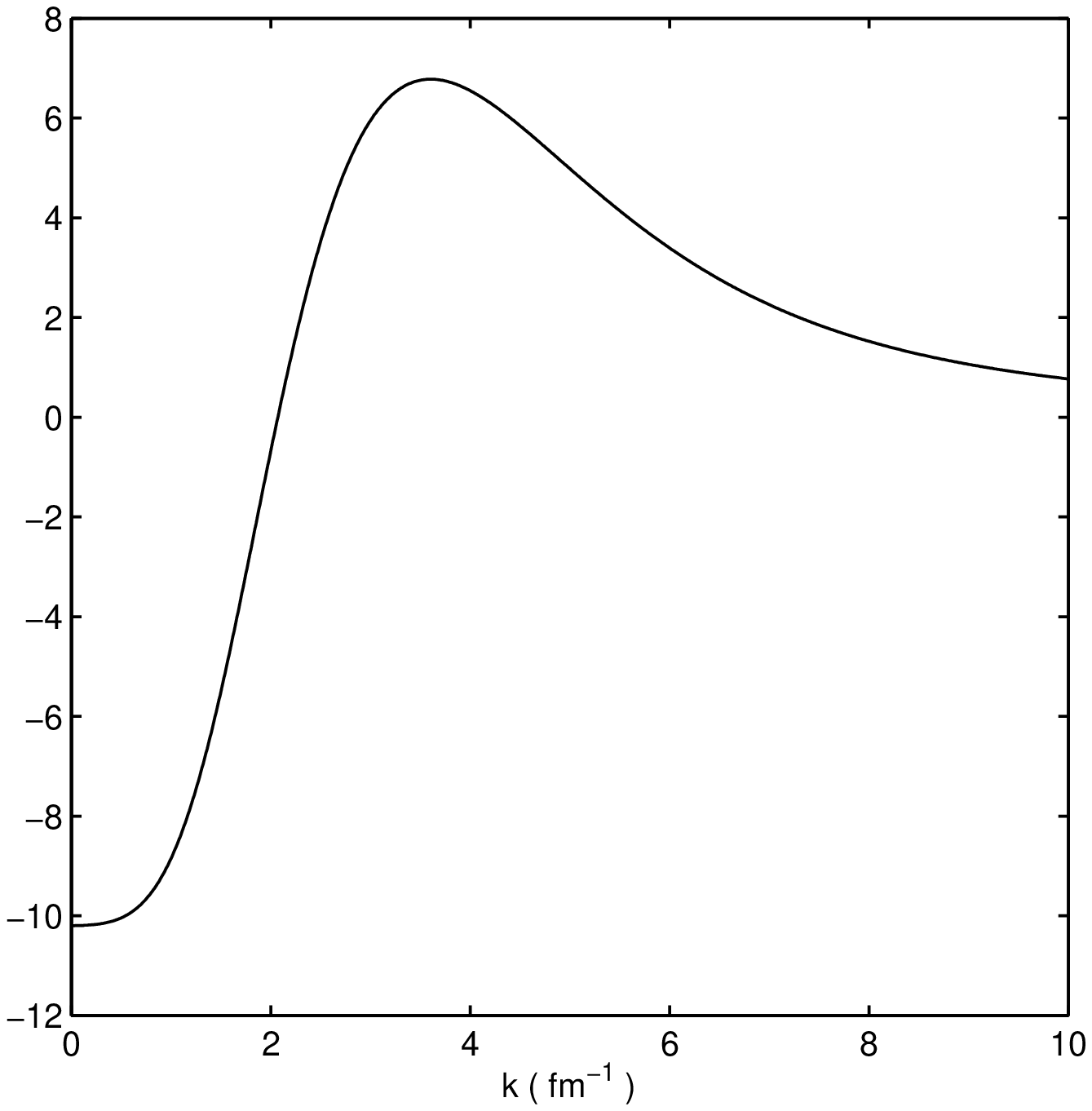}

\vfill\eject

\noindent{\bf Figure 6: Sparse Kernel for $K=3$, $\epsilon=10^{-6}$}

\includegraphics[5cm,7cm][17.5cm,21cm]{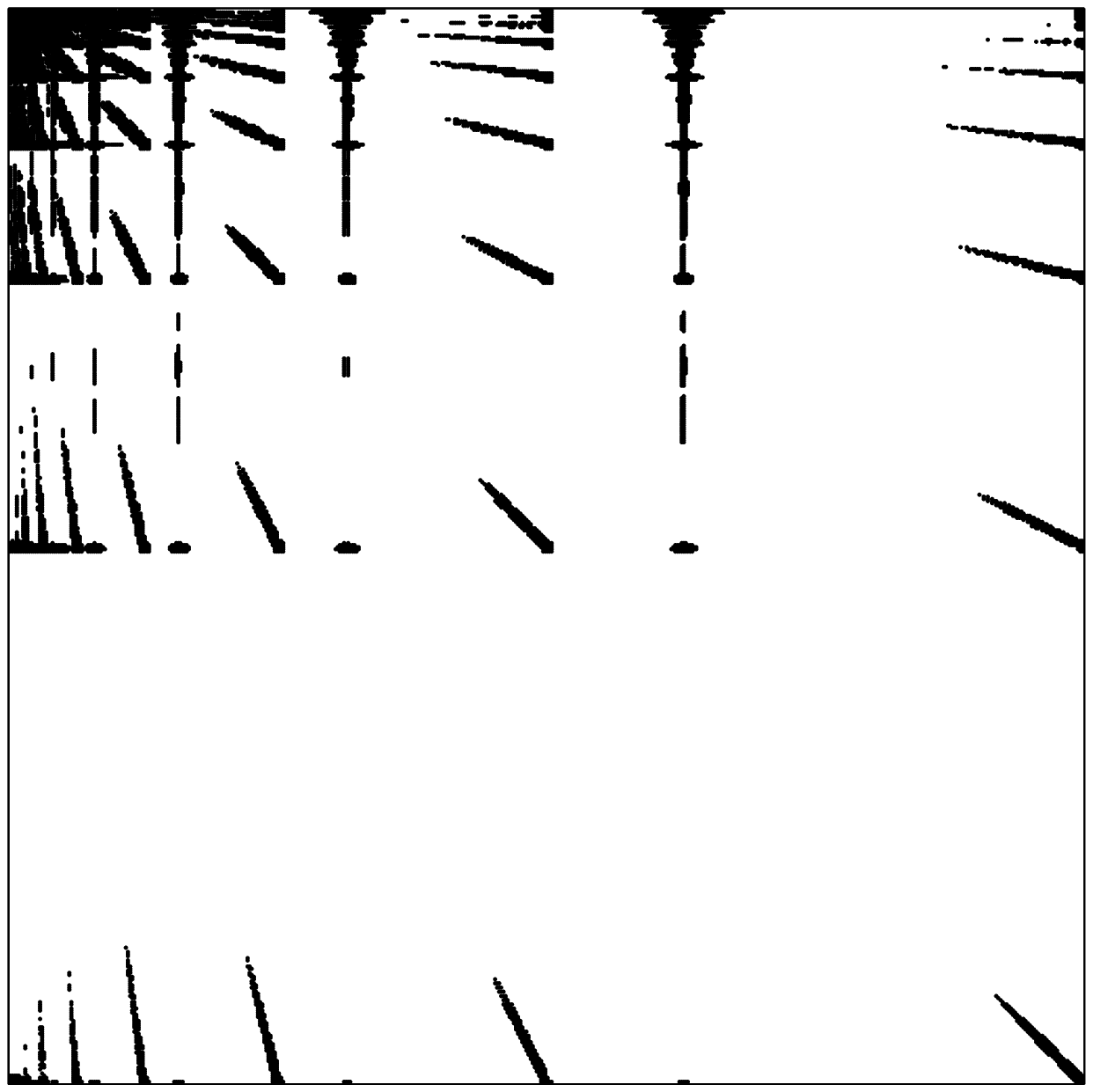}

\end{document}